\documentstyle{natokap}
\begin{opening}
\title{Topological organization of (low-dimensional) chaos.}
\author{Nicholas B. Tufillaro}
\institute{Woods Hole Oceanographic 
Institution \newline
Woods Hole MA\thanks{Current address: CNLS
MS-B258 LANL Los Alamos NM 87545}}    
\date{1 September 1992}
\end{opening}
\begin{document}
\begin{abstract}
Recent progress toward classifying low-dimensional chaos
measured from time series data is described.
This classification theory assigns a template to
the time series once the time series is embedded in three dimensions. The template
describes the primary folding and stretching mechanisms
of phase space responsible for the chaotic motion.
Topological invariants of the unstable periodic orbits
in the closure of the strange set are calculated from 
the (reconstructed) template. These  topological invariants must be 
consistent with any model put forth to describe the
time series data, and are useful in
invalidating (or gaining confidence in) any model
intended to describe the dynamical system generating
the time series.
\end{abstract}
\bigskip

Statistical measures and topological methods are the
two major types of analysis used when studying chaos in smooth
dynamical systems.
These two approaches, the statistical and topological,
often give us different information about the same
dynamical system [Fr].
The ergodic (statistical) theory of dissipative
dynamical systems focuses its attention on an invariant
measure $\mu(\Omega)$ defined on the invariant limit
set $\Omega$ (i.e., a strange attractor or repeller) [Ec].
Information about an invariant measure can have many useful
applications. In time series analysis, for instance, 
$\mu(\Omega)$ is an essential ingredient in building
nonlinear predictive models directly from time series [Ge].

Topological methods of smooth dynamical systems theory are
also of great value in time series analysis.
In particular, in the context of low-dimensional chaos,
topological techniques allow us to develop a classification
theory for chaotic invariant limit sets.
In addition, topological properties often put strong
constraints on the dynamics (for instance, the existence or non-existence
of certain orbits [Ha]). A topological analysis is also an 
essential ingredient for developing rapidly convergent
calculations of the metric properties of the attractor [Cv].

Therefore, when analyzing a time series from a chaotic 
dynamical system we advocate a two step procedure.
First, analyze the topological organization of the
invariant set, and second dress this topological form
with its metric structure. 
We believe, at least in context of low-dimensional chaos,
that as much information as possible should be gleaned from
the topology the chaotic limit set as a first step toward
modeling the dynamics. 
This topological information plays at least two important roles
in applications to time series analysis.
First, topological invariants can be used to identify (or invalidate)
models put forth to explain the data, and second, the topological
classification of chaotic sets serves as a promising first step in
developing predictive models of nonlinear time series data.

Recently, this topological approach to time series analysis has
been worked out in great detail in the context of chaotic
invariant sets of ``low-dimensional'' flows.
In this article, by ``low-dimensional'' we mean flows in
$R^n$ with invariant sets of dimension less than or equal to 3, i.e.,
systems with one unstable direction (one positive Lyapunov exponent).
By restricting our attention to this class of systems, it is 
possible to develop a rather complete physical theory for the
topological classification of such systems and to develop
practical algorithms 
for applying this classification scheme to time series data from 
experiments.
In this article we will review work on this classification
theory. For recent efforts on applying this classification theory
to modeling the dynamics we refer the reader to a
review article by Mindlin and Gilmore [Mi1] which also
contains many practical details about topological time
series analysis. For an elementary introduction to
the knot theory and dynamical systems background 
appropriate for this article see Reference [Tu1].

The major device in this analysis is the template (or knot-holder)
of the hyperbolic chaotic limit set [Ho].
Roughly, a template is an expanding map on a branched surface.
A low-dimensional chaotic limit set with one unstable
direction has a rich set of recurrence properties which
are determined by the unstable saddle periodic orbits
embedded within the strange set. These unstable periodic
orbits provide a sort of skeleton on which the
strange attractor rests. 
For flows in three dimensions, these periodic orbits are closed
curves, or knots. 
The knotting and linking of these periodic orbits
is a bifurcation invariant, and hence these simple 
topological invariants can be used to identify or
``fingerprint'' a strange attractor [Mi2, Tu2].
Templates are central to this analysis
because periodic orbits from a three-dimensional flow of a hyperbolic
dynamical system can be placed on a template in such
a way as to preserve their original topological structure.
Thus templates provide a visualizable model for the topological
organization of the chaotic limit sets. Templates can also
be describe algebraically by finite matrices and this in
turn gives us a quantitative classification theory describing
the primary folding and stretching structure of the strange set [Tu1].

The strategy behind the template theory is as follows.
For a nonlinear dynamical system there are generally
two regimes that are well understood, the regime
where a finite number of periodic orbits exists
and the hyperbolic regime of fully developed chaos.
The essential idea is to reconstruct the form of the
fully developed chaotic limit set from a non-fully developed
(possibly non-hyperbolic) region in parameter space.
Once the hyperbolic limit set is identified, then the
topological information gleaned from the hyperbolic
limit set can be used to make predictions about the
chaotic limit set in other (possibly non-hyperbolic)
parameter regimes, since topological invariants such
as knot types, linking numbers, and relative rotation
rates [So1, So2] are robust under parameter changes.

The identification of a template from a chaotic
time series of low-dimension proceeds in five
steps [Mi3, Mi1]: search for close returns,
three-dimensional embedding of the time series,
calculation of topological invariants, template
identification, and template verification.

In the first step, the search for close returns [Au, Tu2],
the time series is examined for subsegments of the
data which almost return to themselves after n-cycles.
These subsegments of the time series are taken as
surrogates for the unstable (saddle) period-n orbits
which exist in the closure of the strange set.
This search for close returns (unstable periodic orbits)
can be done either before or after the time series 
is embedded in a three-dimensional space [Mi3].

The next step is to embed the time series in
a three-dimensional space. Developing an embedding
procedure which ``optimizes'' the topological
information in the time series is the key to
success with the topological analysis of time
series data. In principal there are several
candidates for an embedding procedure. Both
the method of delays [Pa], and an embedding based
on a singular value decomposition analysis are reasonable choices and are
described by  D. Broomhead in these proceedings.
As a practical matter great care must be taken
to see that the embedding procedure eliminates
any (parametric) drift in the data (for instance,
this may by accomplished by judicious filtering),
and that the embedding procedure  also seeks
to maximize the geometric spatial separation of
the embedded time series trajectory.
With these two criteria in mind, Mindlin and Gilmore [Mi3]
have developed a ``differential phase space embedding''
which works remarkably well for their analysis of
data from the Belousov-Zhabotinskii reaction.
On a case by case basis, finding  
an embedding which ``optimizes'' the extraction of topological
information inherent within the (experimental) time series does
not pose a major obstacle to the analysis. Rather
it suggests that a lot of good work is yet to
be done in developing
a new branch of engineering which might be 
dubbed ``topological signal processing.'' 

In the embedded space, topological invariants
(linking numbers, relative rotation rates, and braid words)
of the surrogate periodic orbits found in the first
step can be calculated. Just a few of these suffice
to determine a template [Mi2, Mi3, Me]. In fact, one can also
identify the template by examining the stretching and
folding of points on the strange attractor as
it evolves through one full cycle [Mc, Le], and also
by examining the ``line-diagram'' of a few geometric
braids calculated from the embedded periodic orbits[Ha]. Thus,
the form of the template is usually very much over determined
by the available experimental data. The fact the
the template is determined from a (small) finite 
amount of information should come as no surprise. Each template
is nothing but a geometric picture for the suspension of a full shift
hyperbolic symbol system which we formally associate
to the (possibly non-hyperbolic) chaotic time series.
This full shift system has the same basic folding
and stretching structure of the original flow, and it
might even be found in the original (experimental) system
in a parameter regime where a chaotic repeller exists.

Once identified, the template can be used to
calculate an additional (infinite) set of
topological invariants including (self) rotation
rates, (self) linking numbers, knot types,
polynomial invariants, and so on. If the 
template identification is correct, these
invariants must all agree with those found
in the time series data. If these invariants
do not agree we can reject the proposed template.
If they all agree, we get added confidence
that the template is correctly identified.
These topological invariants must also agree with
any set of differential equations or other dynamical
model proposed for the data. Thus, this gives us
a way of falsifying (or gaining confidence in)
any proposed model.

Each template itself is equivalent to a 
``framed braid'' [Me]. A framed braid is just
a geometric braid with an integer associate
to each strand called the framing. The linking
of this framed braid is described by a framed braid
linking matrix, and it is this (finite) matrix which
we take as our quantitative (integer) characterization for
the topology of the strange set. For more details 
with an abundance of pictures see Chapter 5 of Reference [Tu1].

The template characterization and classification has recently been
applied to a wide variety of time series data from experimental systems
including the Belousov-Zhabotinskii chemical clock [Mi3], a laser with
a saturable absorber [Pap], an NMR-laser [Tu3], and a ${\rm{CO_2}}$ laser
with modulated losses [Le]. 

The template classification theory is just the beginning of
topological time series analysis. There are many directions
now to take this work. Perhaps the most promising is exploiting the
connection between certain braid types (periodic orbits) and complex behavior
in the flow supporting this braid type. 
Since Thurston's work in the 70's on braid types
and dynamics on the punctured disk, it has been know that
the existence of certain types of braids (i.e., the so called
pseudo-Anosov ones) are sufficient to imply that a dynamical
system has positive topological entropy, that is, 
that the system is chaotic [Th]. Mindlin and Gilmore
found such a braid type (periodic orbit) in their analysis
of the Belousov-Zhabotinskii reaction [Mi3]. It is 
the period-7 pretzel knot of the horseshoe with symbolic name 0110101.
The existence of this single ``non-well ordered orbit'' orbit [Ga]
allows Mindlin and Gilmore to conclude that the system is 
chaotic (at least in the topological sense meaning the existence
of an infinite number of periodic orbits forming a complex
chain recurrent set) without calculating any Lyapunov exponents
or fractal dimensions.

Indeed, as emphasized by D. Broomhead in these proceedings,
some of the most exciting work in nonlinear dynamics is the 
current close interplay between mathematics and experimental
physics. In essence one can seek, in doing an experiment, to show
that certain mathematical hypothesis hold in the given experimental
configuration. If these mathematical hypothesis can
be experimentally verified, then one can learn much more about
the system then either statistical inference or physical experimentation
alone would provide.

{}

\end{document}